\documentclass[prd,aps,preprint,tightenlines,superscriptaddress,nofootinbib]{revtex4-1}

\usepackage{epsfig}
\usepackage{amsmath}

\begin{document}

\title{One-loop Factorization for Inclusive Hadron Production in $pA$
Collisions in the Saturation Formalism}
\author{Giovanni A.  Chirilli}

\affiliation{Nuclear Science Division, Lawrence Berkeley National
Laboratory, Berkeley, CA 94720, USA}
\author{Bo-Wen Xiao}
\affiliation{Department of Physics, Pennsylvania State University,
University Park, PA 16802, USA}

\author{Feng Yuan}
\affiliation{Nuclear Science Division, Lawrence Berkeley National
Laboratory, Berkeley, CA 94720, USA}

\begin{abstract}
We demonstrate the QCD factorization for inclusive hadron production in $pA$
collisions in the saturation formalism at one-loop order, with explicit
calculation of both real and virtual gluon radiation diagrams.
The collinear divergences associated with the incoming parton distribution
of the nucleon and the outgoing fragmentation function of the final state
hadron, as well as the rapidity divergence with small-$x$ dipole gluon
distribution of the nucleus are factorized into the splittings of
the associated parton distribution and fragmentation functions and the
energy evolution of the dipole gluon distribution function. The
hard coefficient function is evaluated at one-loop order, and contains no
divergence.
\end{abstract}

\maketitle

{\it 1. Introduction.}
Gluon saturation at small-$x$ (the longitudinal momentum fraction) in nucleon and nucleus
plays a crucial role in the description of high energy hadronic
scattering~\cite{Gribov:1984tu,Mueller:1985wy,McLerran:1993ni}. It has
been applied to a wide range of processes which are relevant for the
current and future experiments at RHIC and LHC~\cite{arXiv:1002.0333}.
Among these physics, suppression of single inclusive hadron production in the forward direction
of $dAu$ collisions at RHIC has been regarded as one of the most important
evidences for the on-set of gluon saturation at small-$x$ in a large
nucleus~\cite{McLerran:2011zz}.
In particular, the dipole gluon distribution applied to hadron production in
nucleon-nucleus scattering is the same as that in the description of inclusive deep
inelastic scattering structure function at small-$x$~\cite{Dumitru:2002qt}.
The duality between these two processes, and the universality of the gluon saturation is
important to understand the strong interaction dynamics in the dense medium~\cite{Al}.

However, the experimental data are so far interpreted in the leading order
calculation in the saturation formalism~\cite{Albacete:2010bs}. In order to solidate the above
conclusion, we need to go beyond the leading order Born approximation,
and evaluate higher order corrections, which has not yet been done.
A next-to-leading order calculation is highly anticipated in the small-$x$ formalism,
to demonstrate the QCD factorization which can be applied to a wide range
of high energy processes. It has been one of the most important quests in the last few
years~\cite{Gelis:2008rw,Balitsky:2010ze}. Recent studies on the two-particle production in $pA$
collisions have also emphasized the importance of the factorization and
universality issues involved in the hard processes, in particular, for small-$x$
physics~\cite{Dominguez:2010xd}.

In this paper, we will carry out, for the first time, a complete next-to-leading
order (NLO) calculation for the single inclusive hadron production in $pA$
collisions in the saturation regime, by employing both the collinear
factorization and the high energy small-$x$ factorization techniques.
Early efforts have been made in Refs.~\cite{Dumitru:2005gt,Altinoluk:2011qy}.
Our results will not only provide an
important estimate of high order corrections, but also pave the way to
build a QCD factorization for hard processes in the saturation
formalism~\cite{Gelis:2008rw}. Moreover,
this calculation, combining with the associated calculations for inclusive
DIS process~\cite{Balitsky:2010ze}, will help to identify the universality of the dipole gluon
distributions in these processes. The method developed here will be very useful
to other hard processes in small-$x$ physics as well.

In the process of hadron production in $pA$ collisions,
\begin{equation}
p+A\to h+X \ ,\label{pA}
\end{equation}
a parton from the nucleon (with momentum $p$) scatters on the nucleus target
(with momentum $P_A$), and fragments into final state hadron with momentum $%
P_h$. In the dense medium of the large nucleus and at small-$x$, the
multiple interaction become important, and we have to perform the relevant
resummation to make the reliable theoretical calculations. The small-$x$
formalism, the color-glass-condensate or color-dipole formalism, was developed
to calculate these hard processes~\cite{arXiv:1002.0333}. We are particularly interested in the
scattering process with a dilute projectile on a dense target such as
the process of (\ref{pA}), where the
parton from the incoming nucleon can be treated as the normal parton
distribution and it fragments to the final state hadron after multiple
interaction with the nucleus target. According to our following
calculations, the QCD factorization formalism for this process reads as,
\begin{eqnarray}
\frac{d^3\sigma^{ p+A\to h+X}}{dyd^2p_\perp}&=&\sum\limits_a \int \frac{dz}{%
z^2}\frac{dx}{x}
\xi xf_a(x,\mu) D_{h/c}(z,\mu)
\int [dx_\perp] S_{a,c}^Y([x_\perp])\mathcal{H}_{a\to c}(\alpha_s,\xi,[x_\perp]\mu) \ , \label{fac}
\end{eqnarray}
where $\xi=\tau/xz$ with $\tau=p_\perp e^y/\sqrt{s}$, $y$ and $p_\perp$ the
rapidity and transverse momentum for the final state hadron and $s$ the
total center of mass energy square $s=(p+P_A)^2$, respectively. In the above
equation, $f_a(x)$ and $D_{h/c}(z)$ represent the collinear parton distribution from
the incoming nucleon and fragmentation function for the final state hadron,
where $x$ is the momentum fraction of the nucleon carried by the parton $a$,
and $z$ the momentum fraction of parton $c$ carried by the final state
hadron $h$, respectively. The response from the nucleus target is denoted as
$S_{a,c}^Y(x_\perp)$ (see the definitions below), depending on the flavor of the incoming and outgoing partons
and the gluon rapidity $Y$ associated with the nucleus: $Y\approx \ln(1/x_g)$
with $x_g$ being momentum fraction of nucleus entering the hard process.
At the leading order, they are defined as the two-point functions representing the
dipole gluon distribution functions in the elementary and adjoint representations for the
quark and gluon initialed subprocesses~\cite{arXiv:1002.0333}, respectively. Higher order corrections
will have terms that depend on the correlation functions beyond the simple
two-point functions. Because of this reason, the integral $[dx_\perp]$
represents all possible integrals at the particular order. For
example, a four-point function (non-linear term of dipole gluon distribution)
will contribute to the quark channel at one-loop order. The hard factor $\mathcal{H}_{a\to c}$
describes the partonic scattering amplitude of parton $a$ into a parton $c$
in the dense medium. This hard factor includes all order perturbative
corrections, and can be calculated order by order.
Although there is no simple $k_\perp$-factorization form
beyond leading order formalism~\cite{Blaizot:2004wv},
we will find that in the coordinate space, the cross section can
be written into a nice factorization form as Eq.~(\ref{fac}).
Besides the explicit dependence on the variables shown in Eq.~(\ref{fac}),
there are implicit dependences on $p_\perp [x_\perp]$ in the hard
coefficients as well.

The above factorization is derived in the high energy limit,
i.e., $\sqrt{s}\to \infty$, which is the same limit that the dipole formalism
was derived for the inclusive DIS structure function at small-$x$~\cite{CU-TP-441a}.
Two important variables are introduced to separate
different factorizations for the physics involved in this process: the
collinear factorization scale $\mu$ and the energy evolution rapidity
dependence $Y$. The physics associated with $\mu$ follows the normal
collinear QCD factorization, whereas the rapidity factorization $Y$ takes
into account the small-$x$ factorization. The evolution respect to
$\mu$ is controlled by the usual DGLAP evolution, whereas that for
$S_a^Y$ by the Balitsky-Kovchegov (BK) evolution~\cite{Balitsky:1995ub,Kovchegov:1999yj}.
In particular, our one-loop calculations will demonstrate the important
contribution from this rapidity divergence.

A different formula has been proposed in Ref.~\cite%
{Dumitru:2005gt}, where only part of one-loop calculations were taken into account,
and the rapidity divergence is not identified and the collinear evolution effects
are not complete. In particular, the formula in Ref.~\cite{Dumitru:2005gt} was
expressed in the leading order form, which does not allow higher order corrections. Recently, it has
been realized that the higher order corrections are important for high $%
p_\perp$ particle production in $pA$ collisions, referred as ``inelastic"
contribution in Ref.~\cite{Altinoluk:2011qy}. In this paper, we will demonstrate the
factorization formula of Eq.~(\ref{fac}), by a complete one-loop
calculations, including not only the contributions considered in Refs.~\cite%
{Dumitru:2005gt,Altinoluk:2011qy}, but also the virtual diagrams which have not yet been
calculated before. We will take the example of quark channel contribution to
demonstrate this factorization, and list the results for all other channels in the end.
For the quark channel contribution: $qA\to q+X$, the above factorization
formula can be explicitly written as
\begin{eqnarray}
\frac{d^3\sigma ^{p+A\to h+X}}{dyd^2p_\perp}&=& \int \frac{dz}{z^2}\frac{dx}{x%
}\xi xq(x,\mu) D_{h/q}(z,\mu)\int \frac{d^2x_\perp
d^2y_\perp}{\left(2\pi\right)^2} \left\{S^{(2)}(x_\perp,y_\perp)
\left[ \mathcal{H}_{2qq}^{(0)}
+\frac{\alpha_s}{2\pi}\mathcal{H}_{2qq}^{(1)}\right]\right.\nonumber\\
&&\left.+
\int \frac{d^2b_\perp}{(2\pi)^2}
S^{(4)}(x_\perp,b_\perp,y_\perp)\frac{\alpha_s}{2\pi}
\mathcal{H}_{4qq}^{(1)}\right\}\ ,\label{qchannel}
\end{eqnarray}
up to one-loop order.
The two-point and four-point functions are defined as
\begin{eqnarray}
S^{(2)}(x_\perp,y_\perp)&=&\frac{1}{N_c}\langle{U}(x_\perp){U}^\dagger(y_\perp)\rangle_Y\\
S^{(4)}(x_\perp,b_\perp,y_\perp)&=&\frac{1}{N_c^2}\langle \mathrm{Tr}[{U}(x_\perp)%
{U}^\dagger(b_\perp)] \mathrm{Tr}[{U}(b_\perp){U}%
^\dagger(y_\perp)]\rangle_Y\ ,
\end{eqnarray}
with $N_c$ the number of color in QCD, and
$U(x_\perp)=\mathcal{P}\exp\left\{ig_S\int_{-\infty}^{+\infty} \text{d}x^+\,T^cA_c^-(x^+,x_\perp)\right\}$ is the
Wilson line in the small-$x$ formalism~\cite{arXiv:1002.0333} with $A_c^-(x^+,x_\perp)$ being the gluon field solution of the classical Yang-Mills equation. For convenience, we also
introduce the momentum space interpretations of the above
functions:
$\mathcal{F}(k_\perp)=\int\frac{d^2x_\perp d^2y_\perp}{(2\pi)^2}e^{-ik_\perp\cdot
(x_\perp-y_\perp)}S^{(2)}(x_\perp,y_\perp)$,
and $\mathcal{G}(k_\perp,k_{1\perp})=\int\frac{d^2x_\perp d^2y_\perp d^2b_\perp}{(2\pi)^4}%
e^{-ik_\perp\cdot (x_\perp-b_\perp)-ik_{1\perp}\cdot(b_\perp-y_\perp)}
S^{(4)}(x_\perp,b_\perp,y_\perp)$. The above two
functions are related as follows: ${\cal F}(k_\perp)=\int d^2k_{1\perp}{\cal
G}(k_\perp,k_{1\perp})$. At the leading order in $\alpha_s$, there
is only two-point function contribution. At the
next-to-leading order, a non-linear term of the two-point
functions will enter. For the gluon channel, a six-point function
will apply. The goal of this paper is to demonstrate the above
factorization formalism, and obtain the hard coefficients
$\mathcal{H}_{2,4}^{(1)}$. In doing so, we calculate the
scattering amplitude squared at one-loop order, and factorize out
the divergences associated with the splittings of the quark distribution and fragmentation
functions as well as the rapidity divergence from the two-point function.
The hard coefficients are Infra-red and Ultra-violet finite.

{\it 2. One-loop Calculations.}
The leading order results have been calculated before, from which we have
\begin{equation}
\mathcal{H}_{2qq}^{(0)}=e^{-ik_\perp\cdot r_\perp} \delta(1-\xi)\
,
\end{equation}
where $k_\perp=p_\perp/z$ and $r_\perp=x_\perp-y_\perp$.
In the following, we will perform the NLO calculations.
There are virtual and real gluon radiations. We plot the typical diagrams for
them in Fig.~1. After some algebra, we find that the sum of the virtual
diagrams,
\begin{eqnarray}
-&&\frac{\alpha _{s}}{2\pi ^{2}}\int_{0}^{1}d\xi \frac{1+\xi ^{2}}{1-\xi }%
\left\{ C_{F}\int d^{2}q_{\perp }{\cal I} (q_\perp,k_\perp)
+\frac{N_{c}}{2}\int d^{2}q_{\perp }d^{2}{k_{g1\perp }}{\cal J}(q_\perp,k_\perp,k_{g1\perp})  \right\}\ , \label{virtual}
\end{eqnarray}%
where $C_F=(N_c^2-1)/2N_c$, and ${\cal I}$ and ${\cal J}$ are defined as
\begin{eqnarray}
{\cal I}(q_\perp,k_\perp)&=&\mathcal{F}(k_{\perp })\left[ \frac{q_\perp-k_\perp}{(
q_{\perp }-k_\perp)^{2}}-\frac{q_\perp-\xi k_\perp}{%
(q_{\perp }-\xi k_{\perp })^{2}}\right]^2, \nonumber\\
{\cal J}(q_\perp,k_\perp,k_{g1\perp})&=&\left[
\mathcal{F}(k_{\perp })\delta ^{\left( 2\right) }\left( k_{g1\perp
}-k_{\perp }\right) -\mathcal{G}(k_{\perp },k_{g1\perp })\right]\frac{2(q_{\perp }-\xi k_{\perp })\cdot (q_{\perp }-k_{g1\perp })}{(q_{\perp }-\xi
k_{\perp })^{2}(q_{\perp }-k_{g1\perp })^{2}} \nonumber\ .
\end{eqnarray}
In the above results, we notice that the Ultra-violet divergences cancel
out among the virtual diagrams. This can be
seen from the large $q_{\perp }$ behavior of the above integral and the
identity relating ${\cal G}$ and ${\cal F}$. At
large $q_\perp$ limit, the gluonic interaction with nucleus target is not affected by
the medium effect, and the Ward identity applies, therefore the UV divergence
cancels out between the self-energy and vertex diagrams.
The real gluon radiation diagrams have been considered before~\cite{hep-ph/0405266,Dominguez:2010xd}.
In order to obtain the single inclusive particle production cross sections,
we need to integrate out the phase space of the radiated gluon. In the
high energy limit, this integration will collapse the quadrupole into
dipole and non-linear term of dipoles. At the end of the day, we can cast the real contribution into
\begin{eqnarray}
&&\frac{\alpha _{s}}{2\pi ^{2}}\int \frac{dz}{z^{2}}D(z)\int_{\tau
/z}^{1}d\xi \frac{1+\xi ^{2}}{1-\xi }xq(x)\left\{ C_{F}\int d^{2}k_{g\perp }{\cal I}(k_\perp,k_{g\perp})\right.\nonumber\\
&&\left.
~~~+\frac{N_{c}}{2}\int d^{2}k_{g\perp }d^{2}{k_{g1\perp }}{\cal J}(k_\perp,k_{g\perp},k_{g1\perp})\right\} \ ,
\label{real}
\end{eqnarray}%
where $x=\tau /z\xi $.

There are
rapidity divergences in the above results (\ref{virtual},\ref{real}) when
$\xi \rightarrow 1$.  The rapidity divergence
appears when the longitudinal momentum of the
gluon $k^+_g=(1-\xi)p^+$ goes to zero, namely when the rapidity of
 the gluon goes to $-\infty$.
It is important to note that the rapidity
divergence cancels out for the term proportional to the color-factor $C_{F}$.
The remainder of the rapidity divergence is proportional to color-factor $N_{c}/2$,
which is exactly the same as that in the BK evolution equation~\cite{Balitsky:1995ub,Kovchegov:1999yj}.
It is also interesting to note that
the rapidity divergence disappears when one integrates over the
transverse momentum $k_{\perp }$~\cite{Collins:2011zzd}.

\begin{figure}[tbp]
\begin{center}
\includegraphics[width=10cm]{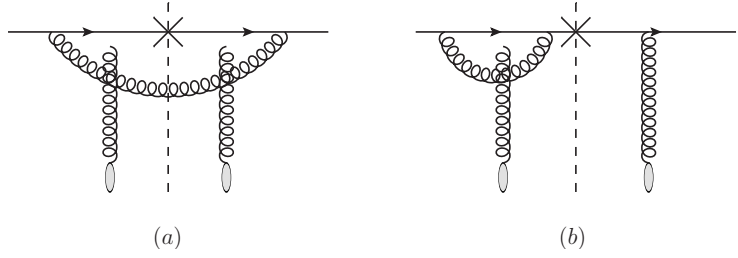}
\end{center}
\caption[*]{Typical Feynman diagrams for the real (a) and virtual (b)
gluon radiation contributions to the quark production $q A\to q+X$  at one-loop order.
The crosses represent the final observed quark, and the blobs in the lower parts of the
diagrams represent the multiple interactions with the nucleus at small-$x$. }
\label{vr}
\end{figure}

There are also collinear divergences from both real and virtual
diagrams. We use dimensional regularization ($D=4-2\epsilon$) and follow the
$\overline{\mathrm{MS}}$ subtraction. Since the
soft gluon radiation has been included into the BK evolution of the
unintegrated gluon distribution, there is no soft divergence.
For example, for the virtual diagrams we have
\begin{eqnarray}
&&\frac{2}{4\pi }\left\{ C_{F}\mathcal{F}(k_{\perp })\left[ -\frac{1}{\epsilon
}+\ln \frac{k_{\perp}^{2}}{\mu ^{2}}\right] +\left( C_{F}-\frac{N_{c}}{2}\right)
\mathcal{F}(k_{\perp })\ln (1-\xi )^{2}\right.\nonumber\\
&&~~+\left.\frac{N_{c}}{2}\int d^{2}k_{g1\perp }\mathcal{G}(k_{\perp },k_{g1\perp })\ln \frac{%
(k_{g1\perp }-\xi k_{\perp })^{2}}{k_{\perp}^{2}}\right\} \ .
\end{eqnarray}%
The collinear divergence is represented by $1/\epsilon$ in the above
equation. The same divergence appears in the real diagram calculations.
By summing up both real and virtual contributions, we identify
the following divergences,
\begin{eqnarray}
&&-\frac{\alpha_{s}N_{c}}{2\pi ^{2}}\int_0^1 \frac{d\xi^{\prime}}{1-\xi^{\prime}}%
\int \frac{d^{2}b_{\perp }}{(2\pi )^{2}}%
e^{-ik_{\perp }\cdot r_\perp}\frac{(x_{\perp }-y_{\perp })^{2}%
}{(x_{\perp }-b_{\perp })^{2}(y_{\perp }-b_{\perp })^{2}} \left[
S^{(2)}(x_{\perp },y_{\perp })-S^{(4)}(x_{\perp },b_{\perp},y_{\perp })\right]  \notag \\
&& + \frac{\alpha_sC_F}{%
2\pi}\int_{\tau/z}^1 d\xi \left(-\frac{1}{\epsilon}\right) \left[\mathcal{P}%
_{qq}(\xi) e^{-ik_\perp\cdot
r_\perp}+\mathcal{P}_{qq}(\xi)\frac{1}{\xi^2}e^{-i\frac{k_\perp}{\xi}\cdot
r_\perp} \right]\frac{1}{(2\pi )^{2}}S^{(2)}(x_{\perp },y_{\perp
})  \ ,\label{total}
\end{eqnarray}
where the splitting kernel is defined as
$\mathcal{P}_{qq}(\xi )=\left( \frac{1+\xi ^{2}}{1-\xi }\right) _{+}$.
Obviously, Eq.~(\ref{total}) contains three divergences:
rapidity divergence, and two collinear divergences. The rapidity divergence
(the first term) can be absorbed into the renormalization of the dipole
gluon distribution~\cite{Balitsky:1995ub,Kovchegov:1999yj,Gelis:2008rw,Balitsky:2010ze}. By doing so,
we introduce the $Y$ dependence in the two-point function, from which
the BK evolution can be understood by identifying $dY= {d\xi' }/{(1-\xi') }$.
The collinear divergences symboled by
$1/\epsilon$ in dimensional regularization will be absorbed into the
renormalization of the quark distribution and fragmentation functions,
following the usual collinear factorization.
After subtracting the above divergences, we obtain the hard
coefficients
\begin{eqnarray}
\mathcal{H}_{2qq}^{(1)} &=& C_{F}\mathcal{P}_{qq}(\xi )\ln
\frac{c_0^2}{r_\perp^2\mu ^{2}}\left(e^{-ik_\perp\cdot
r_\perp}+\frac{1}{\xi ^{2}}e^{-i\frac{k_\perp}{\xi}\cdot
r_\perp}\right)-3C_F\delta(1-\xi)e^{-ik_\perp\cdot r_\perp}\ln
\frac{c_0^2}{r_\perp^2k^{2}_\perp} \notag \ \\
&&-\left( 2C_{F}-N_{c}\right)e^{-ik_\perp\cdot r_\perp}  \left[
\frac{1+\xi^{2}}{\left(1-\xi \right) _{+}}%
\widetilde{I}_{21}-\left( \frac{\left( 1+\xi ^{2}\right) \ln
\left(
1-\xi \right) ^{2}}{1-\xi }\right) _{+}\right]   , \label{h2qq}\\
\mathcal{H}_{4qq}^{(1)}&=&-4\pi N_{c}e^{-ik_\perp\cdot r_\perp}\left\{
e^{-i\frac{1-\xi}{\xi}k_\perp\cdot (x_\perp-b_{\perp})}
 \frac{1+\xi ^{2}}{\left( 1-\xi \right) _{+}}\frac{1}{\xi }\frac{x_\perp-b_\perp
}{\left( x_\perp-b_\perp\right) ^{2}}\cdot \frac{y_\perp-b_\perp}{\left( y_\perp-b_\perp\right) ^{2}}
-\delta (1-\xi )\right. \nonumber \\
&& \left.\times
\int_{0}^{1}d\xi ^{\prime }\frac{1+\xi ^{\prime 2}}{%
\left( 1-\xi ^{\prime }\right) _{+}}\left[\frac{e^{-i(1-\xi')k_\perp\cdot
(y_\perp-b_\perp)}}{(b_\perp-y_\perp)^2}-\delta^{(2)}(b_\perp-y_\perp)\int
d^2 r_\perp' \frac{e^{ik_\perp\cdot r_\perp'}}{r_\perp^{\prime
2}}\right]\right\},
\end{eqnarray}%
where $c_0=2e^{-\gamma_E}$ with $\gamma_E$ the Euler constant, and
\begin{equation}
\widetilde{I}_{21}=\int \frac{d^{2}b_{\perp }}{\pi }\left\{
e^{-i\left( 1-\xi \right) k_{\perp
}\cdot b_{\perp }}\left[ \frac{%
b_{\perp }\cdot \left( \xi b_{\perp }-r_{\perp }\right) }{%
b_{\perp }^{2}\left( \xi b_{\perp }-r_{\perp }\right) ^{2}}-\frac{%
1}{b_{\perp }^{2}}\right] +e^{-ik_\perp\cdot
b_\perp}\frac{1}{b_\perp^2}\right\} \ .
\end{equation}
These hard coefficients do not contain any divergence.
The calculations for all other partonic channels
follow the same procedure, and the
hard coefficients are calculated up to one-loop order.

The above results clearly demonstrate that
the factorization of (\ref{qchannel}) is achieved.
The collinear divergences associated with
the collinear parton distribution and fragmentation functions are
factorized, and so does the rapidity divergence associated with
the two-point function of the nucleus. This is a very
important step to prove the factorization beyond the leading order
in perturbation theory.

Our results also show that we can write down the differential cross section
in a factorization form in the coordinate space. The factorization
scale dependence in the hard coefficients reflect the usual
DGLAP evolutions for the quark distribution and fragmentation
functions. It is interesting to note that similar $\mu$ dependence
(associated with $r_\perp$) has also been found in the transverse
momentum resummation formalism derived for the Drell-Yan
lepton pair production in Ref.~\cite{CERN-TH-3923}. However,
the hard coefficients in our case do
not contain double logarithms, and there is no need for the
Sudakov resummation for inclusive hadron production in $pA$
collisions.

Great simplification for the above hard coefficients can be
found if we take large $N_c$ limit. For example,
the last term of Eq.~(\ref{h2qq}) will drop out.
This applies to all other quark involved channels as well.
Here, we list these hard coefficients in the large $N_c$
limit,
\begin{eqnarray}
H_{2qg}^{(0,1)}&=&H_{2gq}^{(0,1)}=0\ , ~~
H_{2qq}^{(1)}=-\frac{3}{2}N_c\delta(1-\xi)e^{-ik_\perp\cdot r_\perp}\ln
\frac{c_0^2}{r_\perp^2k^{2}_\perp} \ ,~~
\nonumber\\
H_{4qg}^{(1)}&=&-4\pi N_c
\frac{1+(1-\xi)^2}{\xi^2} \frac{x-y}{\left( x_\perp-y_\perp\right) ^{2}}\cdot \frac{b_\perp-y_\perp}{%
\left( b_\perp-y_\perp\right) ^{2}}{\cal W}(\frac{k_\perp}{\xi},k_\perp)\ , \notag \\
H_{4gq}^{(1)}&=&-4\pi
\frac{\xi^2+(1-\xi)^2}{\xi} \frac{x_\perp-y_\perp}{\left( x_\perp-y_\perp\right) ^{2}}\cdot \frac{b_\perp-y_\perp}{%
\left( b_\perp-y_\perp\right) ^{2}}{\cal W}(k_\perp,\frac{k_\perp}{\xi})\ ,
\end{eqnarray}
where ${\cal W}(k_{1\perp},k_{2\perp})=e^{-ik_{1\perp
}\cdot(x_\perp-y_\perp)-ik_{2\perp }\cdot(y_\perp-b_\perp)}$, and
we have chosen $\mu=c_0/r_\perp$ for the factorization scale to
further simplify the above expressions. The complete $\mu$
dependence can be re-stored by using the DGLAP evolution for the
relevant parton distributions and fragmentation functions. With the McLerran-Venugopalan
model for the two-point function~\cite{McLerran:1993ni}, we find
that the best choice for $\mu$ will be in the order of the
saturation scale $Q_s$. For the gluon channel $gA\to gX$, a
factorization similar to Eq.~(\ref{qchannel}) was established,
where the two-point function in the adjoint representation (at
large $N_c$ limit as $S^{(2)}(x,y)S^{(2)}(y,x)$) appears at the LO
and NLO, and a non-linear term as $S^{(4)}(x,b,y) S^{(2)}(y,x)$
enters at NLO. The relevant hard coefficients at large $N_c$ limit
read as: $H_{2gg}^{(0)}=H_{2qq}^{(0)}$,
$H_{2gg}^{(1)}=\frac{22}{9}H_{2qq}^{(1)}$, and
\begin{eqnarray}
H_{6gg}^{(1)}&=&-16\pi N_{c}e^{-ik_{\perp }\cdot r_\perp}\left\{e^{-i\frac{k_{\perp }}{\xi}\cdot
(y-b)} \frac{\left[ 1-\xi (1-\xi )\right] ^{2}}{\left( 1-\xi \right) _{+}}%
\frac{1}{\xi ^{2}}\frac{x_\perp-y_\perp}{\left( x_\perp-y_\perp\right) ^{2}}\cdot \frac{b_\perp-y_\perp}{\left(
b_\perp-y_\perp\right) ^{2}}\right.\notag\\
&&-\left. \delta (1-\xi )\int_{0}^{1}d\xi ^{\prime }\left[\frac{\xi ^{\prime }%
}{\left( 1-\xi ^{\prime }\right) _{+}}+\frac{1}{2}\xi ^{\prime }(1-\xi
^{\prime })\right] \right.\nonumber\\
&&\left.\times \left[\frac{e^{-i\xi' k_\perp\cdot
(y_\perp-b_\perp)}}{(b_\perp-y_\perp)^2}-\delta^{(2)}(b_\perp-y_\perp)\int
d^2 r_\perp' \frac{e^{ik_\perp\cdot r_\perp'}}{r_\perp^{\prime
2}}\right]\right\}\ ,
\end{eqnarray}
respectively. The quark loop contribution has been calculated but not included here.

The above results can be compared to the complete collinear factorization
calculation for inclusive hadron production at large
transverse momentum. This corresponds to the situation
discussed in Ref.~\cite{Altinoluk:2011qy}. Another important issue
is the running coupling effects (see e.g., Ref.~\cite{Horowitz:2010yg}) at one-loop order, which has not been
discussed in this paper. We will carry out
these studies, together with the phenomenological
applications to the RHIC and LHC experiments,
in a separate publication.

{\it 3. Summary.}
We have demonstrated the QCD factorization for inclusive
hadron production in $pA$ collisions in the saturation formalism.
The collinear divergences are shown to be factorized into
the splittings of the parton distribution from the incoming nucleon and the
fragmentation function for the final state hadron. The rapidity divergence
at one-loop order is factorized into the BK evolution for the dipole gluon
distribution of the nucleus. The hard coefficients are calculated up to
one-loop order. In principle, using these hard coefficients together with the NLO parton distributions and fragmentation functions as well as the NLO small-$x$ 
evolution equation\cite{Kovchegov:2006vj,Balitsky:2008zza}, one can obtain
the complete NLO cross section of the inclusive hadron production in $pA$ collisions. 

These results are very important, not only for the phenomenological applications
to the inclusive hadron production in $pA$ collisions at RHIC and the LHC
where active experiments are pursued for the study of saturation physics,
but also for theoretically promoting the rigorous developments toward a
complete QCD factorization in small-$x$ physics~\cite{Gelis:2008rw}. Our results
will stimulate further applications of the method, in particular, the factorization technique,
used in this paper to other hard processes involving big nucleus and small-$x$
gluon distributions, as well as those in hot/dense medium. We expect more
exciting developments along this line.

We thank I. Balitsky, F. Dominguez, F. Gelis, J. Jalilian-Marian, C. Marquet, A. H. Mueller, J.W. Qiu, A. Stasto, and R. Venugopalan
for discussions and comments. This work was supported in
part by the U.S. Department of Energy under the contracts DE-AC02-05CH11231
and DOE OJI grant No. DE - SC0002145.

\end{document}